\documentclass[reprint,
amsmath,
amssymb,
aps,
pra,
]{revtex4-2}

\usepackage{amsthm} 
\usepackage{booktabs} 
\usepackage{color} 
\usepackage{mathpazo} 
\usepackage{mathrsfs} 
\usepackage{slashed} 
\usepackage{tikz,tikz-cd} 
\usepackage{comment} 

\renewcommand{\d}{\mathrm{d}}
\newcommand{\dbar}{\bar{\partial}}

\newcommand{\p}{\partial}

\newcommand{\be}{\begin{equation}}
\newcommand{\ee}{\end{equation}}

\newcommand{\bea}{\begin{equation}\begin{aligned}}
\newcommand{\eea}{\end{aligned}\end{equation}}

\newcommand{\half}{\tfrac{1}{2}}

\newcommand{\op}{\operatorname}

\newtheorem{theorem}{Theorem}

\newcommand{\abs}[1]{\left| #1 \right|}
\newcommand{\ip}[1]{\left\langle #1 \right\rangle}

\newcommand{\norm}[1]{\left\| #1 \right\|}

\newcommand{\tri}{\triangle}
\newcommand{\what}{\widehat}

\newcommand{\la}{\langle}
\newcommand{\ra}{\rangle}

\newcommand{\C}{\mathbb{C}}
\newcommand{\R}{\mathbb{R}}

\newcommand{\CP}{\mathbb{CP}}
\newcommand{\PT}{\mathbb{PT}}

\newcommand{\mc}{\mathcal}

\newcommand{\im}{\mrm{i}}
\newcommand{\cA}{\mathcal{A}}
\newcommand{\cB}{\mathcal{B}}

\newcommand{\cM}{\mathcal{M}}

\newcommand{\mf}{\mathfrak}

\newcommand{\fg}{\mathfrak{g}}

\newcommand{\fsl}{\mathfrak{sl}}

\newcommand{\mrm}[1]{\mathrm{#1}}

\renewcommand{\cha}{\mathrm{ch}}

\renewcommand{\t}{\mathrm{t}}
\newcommand{\todd}{\mathrm{td}}

\newcommand{\Cha}{\mathrm{Ch}}

\newcommand{\gSO}{\mathrm{SO}}

\newcommand{\gSU}{\mathrm{SU}}
\newcommand{\gU}{\mathrm{U}}

\newcommand{\Oo}{\mathscr{O}}

\newcommand{\al}{\alpha}
\newcommand{\eps}{\epsilon}

\newcommand{\da}{{\dot\alpha}}
\newcommand{\db}{{\dot\beta}}

\newcommand{\lt}{{\tilde\lambda}}

\newcommand{\sfa}{\mathsf{a}}

\begin{document}

\title{Scattering in Instanton Backgrounds}

\author{Roland Bittleston}
 \email{rbittleston@perimeterinstitute.ca}
\author{Kevin Costello}
\email{kcostello@perimeterinstitute.ca}
\affiliation{Perimeter Institute for Theoretical Physics,\\ 31 Caroline Street, Waterloo, Ontario, Canada}

\date{\today}

\begin{abstract}
In this letter we evaluate one-loop all-plus gluon amplitudes of $\gSU(N_c)$ gauge theory with $N_f$ fundamental fermions in the presence of a flavour instanton background.  Fermion zero modes are regulated with a chiral mass term.  This computation is performed by cancelling a twistorial 't Hooft anomaly via the Green-Schwarz mechanism.  We find that the trace-ordered amplitude has the form of a Parke-Taylor factor multiplied by the Fourier transform of the instanton density evaluated on the total momentum of the gluons.  A background flavour instanton modifies the leading soft gluon and photon theorem, generating a level equal to twice the instanton charge in the soft Kac-Moody symmetry.  We discuss the implications of our results for amplitudes in the presence of dynamical instantons.
\end{abstract}

\maketitle


\section{Introduction}

An exact formula for the one-loop scattering of an arbitrary number of positive helicity gluons has been known since the classic works of \cite{Bern:1993qk,Bern:1993sx,Mahlon:1993si}.  In this paper we determine an exact formula for this scattering process for $\gSU(N_c)$ gauge theory with $N_f$ fundamental fermions, in the presence of an arbitrary self-dual background gauge field for the $\gSU(N_f)$ flavour symmetry.  We will perform this computation in self-dual gauge theory where the only Feynman diagrams contributing at any loop order are such that each connected component has exactly one loop.  Only the connected one-loop amplitude coincides with that of the non-self-dual gauge theory.

The presence of fermion zero modes makes amplitudes in an instanton background a little subtle.  Perhaps the simplest way to interpret our computation is that we are calculating the effective measure on the space of fermion zero modes.  Alternatively, as we will explain in more detail shortly, we can interpret our computation as giving the actual amplitude where the fermion zero modes have been lifted by introducing a chiral mass term $m\eps^{\db\da}\tilde\psi_\da\tilde\psi_\db$ \footnote{We could also resolve the zero mode issue by simultaneously scattering $2kN_c$ right-handed fermions.  Tree level amplitudes of this type have previously been investigated in \cite{HawkingYM,Ringwald:1989ee}.}.

Our formula is very simple.  Let $A_f$ denote a background gauge field for the $\gSU(N_f)$ flavour symmetry group.  We assume that $F_-(A_f) = 0$. Let
\be \mc{D}(A_f,x) = - \frac{1}{8\pi^2}\op{tr}(F(A_f)^2)(x) \ee
be the instanton density, normalized so that the integral over $x$ of $\mc{D}(A_f,x)$ is the instanton number, which is non-negative for self-dual instantons. Let $\what{\mc{D}}(A_f,p)$ denote its Fourier transform.

We are interested in the scattering in self-dual $\gSU(N_c)$ gauge theory with $N_f$ Diracs, in the presence of a flavour symmetry instanton.  The only non-trivial connected scattering process in self-dual gauge theory is the one-loop all-plus amplitude.  We will consider both the connected trace-ordered partial amplitude, and the disconnected amplitude.

The connected, trace-ordered amplitude will be denoted
\be \mc{A}^\textup{trace-ordered}(1^+, \dots, n^+; A_f) \ee
and full amplitude will be denoted
\be \mc{A}^\textup{disconnected}(1^+,\dots,n^+; A_f)\,. \ee
Then:
\begin{theorem} \label{thm:instanton_amplitudes}
\be \label{eqn:main} \mc{A}^\textup{trace-ordered}(1^+,\dots,n^+;A_f)  
= - 2\frac{\what{\mc{D}}(A_f,P)}{\ip{12}\dots \ip{n1}} \ee
where $P$ is the total momentum of the $n$ incoming gluons. 
\end{theorem}
The disconnected amplitude is a sum of products of connected amplitudes, in the usual way.

The same formula holds, multiplied by a factor of $-\half$, if we scatter $N_s$ complex scalars charged under a background $\gSU(N_s)$ instanton.  


\section{Self-dual Gauge Theory and Twistor Space} 

This result is proved by studying self-dual gauge theory and its uplift to twistor space. 

Let us briefly review the self-dual limit of gauge theory \cite{Chalmers:1996rq}.  The dynamical fields of self-dual Yang-Mills are a gauge field $A$ and an adjoint valued anti-self-dual $2$-form $B$. We may further couple to Weyl fermions $\psi_-,\tilde\psi_+$ in the representation $R_f$, and complex scalars $\phi,\tilde\phi$ in the representation $R_s$. The Lagrangian is
\be \label{eqn:sdgaugetheory} \int_{\R^4}\op{tr}(B \wedge F_-(A) ) + \la\psi_-,\slashed{D}_A\tilde\psi_+\ra_{R_f} + \la\phi,\triangle_A\tilde\phi\ra_{R_s}\,. \ee
Full perturbative gauge theory is recovered by adding the term $\half g^2_\text{YM}\op{tr}(B^2)$.  The one-loop all-plus amplitude is insensitive to the deformation term $\op{tr}(B^2)$, and so can be computed in the self-dual theory; in fact it is the only amplitude of the self-dual theory \footnote{At least for generic external kinematics.  There are non-vanishing distributional terms \cite{Guevara:2026qzd}.}.

Classically, self-dual Yang-Mills theory arises from a field theory on twistor space called holomorphic BF theory \cite{Ward:1977ta,Mason:2005zm, Boels:2006ir}.  At loop-level, this generally fails to be true \cite{Costello:2021bah}; there is a one-loop gauge anomaly coming from a box diagram on twistor space.  This anomaly can be removed by adding a counter-term on twistor space which is non-local, but which is local on space-time.  From the space-time perspective, this is an anomaly to integrability \cite{Bardeen:1995gk}.

A quantum field theory on space-time which arises from an anomaly free holomorphic theory on twistor space was called a \emph{twistorial} theory in \cite{Costello:2021bah}.  Twistorial theories have no scattering \cite{Costello:2022wso}, which is why we can infer information about scattering from anomalies on twistor space.  

Anomalies on twistor space can sometimes be cancelled by a Green-Schwarz mechanism \cite{Costello:2021bah} . On space-time, the Green-Schwarz mechanism involves introducing a fourth-order scalar field $\rho$, with kinetic term $\rho \tri^2 \rho$ (where $\tri$ is the Laplace operator).  Twistor anomalies, i.e., anomalies to integrability, are cancelled by coupling $\rho$ to the topological charge $\op{tr}(F(A)^2)$.  This method successfully reproduces \cite{Dixon:2024mzh} the standard one-loop all-plus amplitude.

In our context, we want to cancel the mixed gauge-flavour anomaly and so introduce a term proportional to $\rho(\op{tr}(F(A)^2 ) - \op{tr}(F(A_f)^2  ))$.  We can view the $\rho\op{tr}(F(A_f)^2)$ term as a source for $\rho$, since $A_f$ is not dynamical.

Anomaly cancellation implies that the one-loop amplitude coming from a loop of fermions in the presence of a flavour gauge field $A_f$ is cancelled by the tree-level diagram coming from gluons scattering of the axion field $\rho$ sourced by $\op{tr}(F(A_f)^2)$.  As we explain in the supplementary material \ref{sec:thm_proof}, this analysis leads to the formula \eqref{eqn:main}.


\section{Instanton Corrections to the Soft Gluon and Photon Theorems}

The expression \eqref{eqn:main} is closely related to the Kac-Moody correlator at level $2k$.  Let
\be \ip{J_{\sfa_1}(z_1) \dots J_{\sfa_n}(z_n)}_k \ee
be the level $k$ Kac-Moody correlator, where $\sfa_1,\dots,\sfa_n$ are adjoint indices.  Then, formula \eqref{eqn:main} implies that \footnote{Here the disconnected amplitude does not include diagrams with loops of $\gSU(N_c)$ gluons.  Their contribution is given by the standard formula \cite{Bern:1993sx}.}
\be \label{eqn:disconnected}
\lim_{p_i \to 0} \mc{A}^\textup{disconnected}(1^+,\dots,n^+;A_f)  = \ip{J_{\sfa_1}(z_1) \dots J_{\sfa_n}(z_n)}_{2k} \ee
where the spinors $\lambda_i$ built from the null momenta $p_i$ are related to $z_i$ by $\lambda_i = (1,z_i)$.

This result implies a correction to the leading soft gluon theorem in an instanton background. To state the leading soft gluon theorem, consider a general perturbative amplitude $\mc{A}\big((1^+,\t_{\sfa_1}),\dots,(n^\pm,\t_{\sfa_n})\big)$ where we've included the dependence on the colour matrices $\t_{\sfa_i}$ in the notation.  The leading soft gluon theorem \cite{Low:1958sn, Weinberg:1965nx,Bassetto:1983mvz} predicts that
\bea
&\lim_{p_1 \to 0} \mc{A}\big((1^+,\t_{\sfa_1}),(2^{\pm},\t_{\sfa_2}),\dots,(n^\pm,\t_{\sfa_n})\big) \\
&\sim \sum_{j = 2}^n \frac{1}{\ip{1j}} \mc{A}\big((2^\pm, \t_{\sfa_2}),\dots,(j^\pm,[\t_{\sfa_1},\t_{\sfa_j}]),\dots,(n^\pm,\t_{\sfa_n})\big)\,.
\eea
The leading soft gluon theorem is true to all orders in self-dual gauge theory. In full Yang-Mills theory, the leading soft gluon theorem receives loop corrections from IR divergences \cite{Bern:2014oka}.

Our formula for scattering in an instanton background gives a correction to the leading soft gluon theorem, coming from a Feynman diagram which has a connected component which is a scalar loop with two external $\gSU(N_c)$ gluons.  We find that, for the positive helicity amplitudes we are considering, the soft limit has an extra term
\bea
&\lim_{p_1\to 0}\mc{A}\big((1^+,\t_{\sfa_1}),(2^+,\t_{\sfa_2}),\dots,(n^+,\t_{\sfa_n})\big) \\
&\sim 2\sum_j\frac{\what{\mc{D}}(A_f,p_j)}{\ip{1j}^2}\op{tr}(\t_{\sfa_1} \t_{\sfa_j})\mc{A}\big((2^+,\t_{\sfa_2}),\dots, \\
&\hspace{12em}\what{(j^+,\t_{\sfa_j})},\dots, (n^+,\t_{\sfa_n})\big)\,.
\eea
He, Mitra and Strominger \cite{He:2015zea} show that the leading soft gluon theorem corresponds to a level 0 Kac-Moody symmetry.  To understand the commutator between the soft gluon currents, we should consider the amplitude in the presence of two soft gluons, which we take to be states $1$ and $2$.  The commutator is encoded in the poles in $z_1 - z_2$.  When we have an instanton background, as well as the usual soft poles, we have an extra pole of the form
\be 2\frac{\what{\mc{D}}(A_f,0)}{\ip{12}^2}\op{tr}(\t_{\sfa_1}\t_{\sfa_2})\,. \ee
Tautologically, the Fourier transform of the instanton density evaluated at momentum $0$ is the instanton charge $k$. This means that the extra pole we have is $2k/\ip{12}^2$, leading to the following result:
\begin{theorem}
In the presence of an instanton of charge $k$, the leading soft theorem leads to a Kac-Moody symmetry at level $2k$.
\end{theorem}
We can replace $\gSU(N_c)$ by $\gU(1)$, in which case we find a correction to the soft photon theorem. As we will discuss shortly, we can upgrade flavour instantons to dynamical instantons by performing an integral over the moduli space of $\gSU(N_f)$ instantons.  Since the computation we have performed is at the level of the integrand, it remains true once we integrate over the charge $k$ instanton moduli space.


\section{Dynamical Instantons and Zero Modes} 

Now suppose we have $\gSU(N_1)\times\gSU(N_2)$ gauge theory with bifundamental fermions.  We gave a formula for scattering of $\gSU(N_1)$ gluons in the presence of an $\gSU(N_2)$ instanton.  We can make the $\gSU(N_2)$ into dynamical (self-dual) gauge fields by integrating over the moduli of $\gSU(N_2)$ instantons, with the  one-loop determinant measure.  

The one-loop measure on instanton moduli space is of the form \cite{tHooft:1976snw,Vainshtein:1981wh}
\be
\rho^{bk - 5}\d\rho\,\d^4X\,\d\!\op{Vol}_{\mc{M}^0_k}
\ee
where $X$ are centre of mass coordinates, $\rho$ is the size of the instanton, $\mc{M}^0_k$ is the compact moduli space of instantons of fixed size and position, and
\be
b = \frac{11}{3} N_2 - \frac{2}{3} N_1
\ee
is the $\beta$-function coefficient for the $\gSU(N_2)$ gauge theory with $\gSU(N_1)\times\gSU(N_2)$ bifundamental fermions.  Integrating our amplitude against this measure gives an expression for the contribution of dynamical $\gSU(N_2)$ instantons to the scattering of $\gSU(N_1)$ gluons. 

Before we write the formula more explicitly, let us discuss a very important subtlety.  Fermions in the presence of instantons acquire zero modes.  In the presence of an instanton of charge $k$, the fermions $\tilde\psi_\da$ acquire zero modes which live in $k$ copies of the fundamental plus antifundamental of $\gSU(N_1)$ \footnote{We are treating the space of fermions as complex, not real, vector space, in which the path integral is taken over a contour. This is natural because instantons are Euclidean and Euclidean spinors $\psi_\alpha,\tilde\psi_\da$ are complex fields.}.  Without the insertion of an additional operator, the presence of zero modes would render the amplitude trivial, as the Grassmann integral over the zero modes would yield zero.    

We solve the problem of zero modes in a very simple way.  We introduce a \emph{chiral} mass $m\eps^{\db\da}\tilde\psi_\da\tilde\psi_\db$.  We do not include the anti-chiral component $m\eps^{\beta\alpha}\psi_\alpha\psi_\beta$ of the usual mass term.  Because the spinor zero modes have only dotted spinor indices, this mass term lifts the zero mode degeneracy. 

't Hooft \cite{tHooft:1976snw} showed that the insertion of operators soaking up the zero modes suppresses the UV effects from small instantons.  Let us show how this phenomenon appears in our analysis.  With a chiral mass, the fermion kinetic term is
\be \psi^\alpha D_{\da\al}\tilde\psi^\da + \frac{m}{4}\eps^{\db\da}\tilde\psi_\da\tilde\psi_\db\,. \ee
The classical Lagrangian of self-dual gauge theory is scale invariant as long as we asymmetrically assign the $\tilde\psi_\da$ fermions dimension $2$ and the $\psi_\alpha$ fermions dimension $1$.  This differs from the usual symmetric assignment of dimensions by the axial symmetry $\gU(1)_A$.  Then, we can integrate out the $\tilde\psi_\da$ fermions, leaving only the $\psi_\alpha$ fermions which now acquire the scalar-field kinetic term
\be
\frac{1}{2m}\eps^{\beta\alpha}D_\mu\psi_\alpha D^\mu\psi_\beta\,.
\ee
We have shown that having $N_f$ fundamental fermions with a chiral mass is equivalent to having $2 N_f$ complex fundamental scalars which have Grassmann parity. The Lorentz group action on the scalars is not the standard one, but is twisted by identifying the $\gSU(2)$ rotating the undotted spinor indices with an $\gSU(2)$ internal symmetry inside $\gSU(2N_f)$. 

Turning fermions into wrong-parity scalars in this way changes the $\beta$-function coefficient in the one-loop measure on instanton moduli space. Suppose that as before we have $\gSU(N_1) \times \gSU(N_2)$ gauge theory with bifundamental fermions.  The $\beta$-function coefficient of the $\gSU(N_2)$ coupling is $b = \tfrac{11}{3} N_2 - \tfrac{2}{3} N_1$.  After introducing a chiral mass, the contribution of the fermions becomes that of wrong-parity scalars, and we find a $\beta$-function coefficient
\be
\tilde{b} = \frac{11}{3}N_2 + \frac{1}{3}N_1\,.
\ee
Since fermions with a chiral mass are equivalent to wrong-parity scalars,  it is this modified $\beta$-function coefficient that appears in the measure on instanton moduli space, which is now of the form
\be
\rho^{\tilde{b}k - 5}\d\rho\,\d^4X\,\d\!\op{Vol}_{\mc{M}^0_k}\,.
\ee
This differs by a factor of $\rho^{N_1k}$ from before the introduction of the chiral mass term $m\eps^{\db\da}\psi_\da\psi_\db$.  This analysis does not pick up the overall factor of $m$ which is hidden in the normalization.  Since there are $2N_1k$ zero modes, including the dependence on $m$ gives us the factor $(m\rho)^{N_1k}$ multiplying the amplitude before zero modes were absorbed.

Note that $\tilde{b} > 4$ for all $N_2 \ge 2$, so that as long as we have a theory with no scalars, there is never a UV divergence coming from small instantons. 


\subsection{Explicit Formula for Dynamical Instanton Contribution to Scattering}

Now let us make the formula sketched above more explicit. Suppose our scattering process has $l$ traces, where the $i$\textsuperscript{th} trace has $n_i$ gluons.  We will write down the multi-trace ordered amplitude \footnote{
At the level of the integrand over instanton moduli space, multi-trace amplitudes are products of single-trace amplitudes, but this is no longer true when we integrate.}, which is the coefficient of the colour factor 
\begin{equation}
\op{tr}(\t_{\sfa_1}\dots\t_{\sfa_{n_1}})\dots\op{tr}(\t_{\sfa_{n_1 + \dots + n_{l-1} + 1}}\dots\t_{\sfa_{n_1 + \dots + n_l}})\,.
\end{equation}
Let $P_i$ be the total momentum in the $i$\textsuperscript{th} trace.  The formula for the contribution of dynamical $\gSU(N_2)$ instantons to scattering of $\gSU(N_1)$ gluons is
\bea \label{eqn:amplitudeintegral}
&\mc{A}_{n_1,\dots,n_l} = (-2)^l \int\rho^{\tilde{b}k-5}\d\rho\,\d^4X\, \d\!\op{Vol}_{\mc{M}^0_k} \\
&\op{PT}_1 \dots \op{PT}_l\what{\mc{D}}(A_2,P_1) \dots \what{\mc{D}}(A_2,P_l)\,.
\eea
In this expression $A_2$ is the gauge field of the $\gSU(N_2)$ instanton parametrised by $\rho,X$ and $\cM_k^0$.  $\op{PT}_i$ is the Parke-Taylor factor associated to the $i$\textsuperscript{th} trace, so that for example
\be \op{PT}_1 = \frac{1}{\ip{12}\ip{2 3}\dots\ip{n_1 1}}\,. \ee

In the case of an instanton of charge $1$, the moduli space is parametrised (up to $\gSU(N_2)$ transformations) by only $\rho$ and $X$. In this case explicitly compute the Fourier transformed instanton density:
\be
\widehat{\mc{D}}(a,p) = \frac{1}{2}e^{\im p\cdot X}\rho^2p^2K_2(\rho|p|)
\ee
where $K_2$ is a modified Bessel function of the second kind. 
This gives us a completely explicit formula for the contribution of a single instanton $\gSU(N_2)$ instanton to scattering of $\gSU(N_1)$ gluons. 

The instanton correction to the soft gluon or photon theorem is evident from this expression.  Suppose there are only two gluons with momenta $p_1$ and $p_2$ in the first trace.  The first then trace contributes a factor of 
\be \frac{1}{\ip{12}^2}2\rho^2 p_1\cdot p_2K_2(\rho\sqrt{2p_1\cdot p_2})\,. \ee
We can take the limit $p_1 \to 0$ while assuming that $p_1 \cdot p_2$ is real and non-negative.  In this limit, the contribution of the first trace becomes $2/\la12\ra^2$, which is the Kac-Moody OPE at level $2$. 

If we use $\gU(1)$ instead of $\gSU(N_1)$, this becomes a correction to the leading soft photon theorem.  Rewriting in terms of polarization tensors $\eps_i$ associated to the momenta $p_i$, subject to the positive-helicity constraint $\eps^{\db\da}\eps_{i,\da\al} p_{i,\db\beta} = 0$, the soft factor is
\be
2\frac{(\eps_1 \cdot p_2)(\eps_2\cdot p_1) - (p_1\cdot p_2)(\eps_1 \cdot \eps_2)}{(p_1\cdot p_2)^2}\,.
\ee
Clearly this has a pole as $p_1 \to 0$, correcting the leading soft gluon (or photon) theorem.  Since the leading soft photon theorem is exact to all orders in perturbation theory, this computation is in tension with the idea that instanton contributions should appear by resumming perturbative expansions.


\subsection{Contribution of Large Instantons}

Suppose the total momentum $P_i$ in each trace is space-like.  In that case, $\abs{P_i} = \sqrt{P_i \cdot P_i}$ is positive, and the Bessel function $K_2(\rho \abs{P_i})$ decays exponentially for large $\rho$.  This means the integral \eqref{eqn:amplitudeintegral} is absolutely convergent.  (Convergence for small $\rho$ is automatic, because $\tilde{b} > 4$.)

In the supplementary material \ref{supp:decay}, we use a formula of Osborn \cite{Osborn:1981yf} to show that this phenomena is true for instantons of any charge: the Fourier transformed instanton density has exponential decay at large real momentum. This implies that the integral \eqref{eqn:amplitudeintegral} is absolutely convergent under the same conditions on the momenta.  

A subtlety is that this argument can only hold for multi-trace amplitudes, because for single-trace amplitudes conservation of momentum must hold. Roughly speaking, large instantons whose centre of mass is far away from the origin can contribute IR divergences to single-trace amplitudes.  There are no IR problems for single-trace form factors, where translation symmetry has been broken by the insertion of a local operator.


\section{UV Divergences in the Presence of Scalars}

We have seen that there are no UV or IR divergences to the instanton amplitude when our  theory has only fermions.  Let us now consider the same set-up as before, but where instead of bifundamental fermions we have bifundamental scalars.  The one-loop $\beta$-function coefficient is now
\be
b = \frac{11}{3} N_2 - \frac{1}{6} N_1\,.
\ee
The integral over instanton moduli space has UV divergences coming from small instantons when $b \le 4$. Let us choose $N_1,N_2$ so that $b = 4$ (for example, $N_2 = 2$ and $N_1 = 20$) and let us introduce a cut-off so that the parameter $\rho$ is integrated over the region $\rho \ge \eps$.

Consider the effect of a small instanton of charge $1$.  The small instanton contribution to the scattering amplitude has a UV divergence of the form
\be
\op{Vol}(\mc{M}^0_1)\log\eps    
\ee
which multiplies the scattering of a size zero instanton.  

The instanton density for a size zero instanton of charge one at $X$ is of course a $\delta$-function, whose Fourier transform is $e^{\im P\cdot X}$. 
As we have seen, scattering off a size zero instanton at $X$ is the Kac-Moody correlator
\be e^{\im P\cdot X} \ip{J_{\sfa_1}(z_1)\dots J_{\sfa_n}(z_n)}_{-1}\,. \ee
Integrating over $X$, we find that the UV divergence from small instantons is
\be \delta^4(P)\log\eps\ip{J_{\sfa_1}(z_1)\dots J_{\sfa_n}(z_n)}_{-1}\,.
\ee
In order to remove this UV divergence, we need to introduce a counter-term given by the Kac-Moody correlator.  Such a counter-term would be the integral of an operator whose form factor is the Kac-Moody correlator.  

There is no such local operator, for the simple reason that it would need to be of dimension $0$.  We conclude that, in the case $b = 4$, the theory has a UV divergence which can not be cancelled by a local counter-term. 

If we studied instantons of charge $k$, there are UV divergences of this type coming from the region in instanton moduli space where the instanton becomes a singular instanton of charge $l\le k$, times a smooth instanton of charge $l-k$.  This singularity will be accompanied by a Kac-Moody correlator at level $-l$.

What this computation tells us is that the non-perturbative completion of self-dual gauge theory has an additional parameter, as well as the self-dual theta angle $\theta_{SD}$; and that this additional parameter flows. This parameter appears in the choice of integration measure in the moduli space of instantons.  To the usual measure on the moduli space of instantons, we can add $\delta$-function measures on moduli of instantons which have singularities. 

The new measure can not be arbitrary, but must factorize as a product when an instanton decomposes into two well-separated instantons of lower charge.  The term discussed above gives a $\delta$-function measure on the locus in the moduli of instantons of charge $k$ which are a product of instantons of charge $l$, with $l-k$ point-like instantons of charge $1$.  There are other possible $\delta$-function measures which factorize for well-separated instantons. For instance, we can add a $\delta$-function measure on the locus in instanton moduli space where two point-like instantons of charge $1$ coincide. We will refer to these $\delta$-function measures as instanton counter-terms.

Instanton counter-terms must transform under scaling in the same way as the measure on the interior of instanton moduli space. The measure on the interior is of dimension $-bk$.  The allowed measures on the locus of point-like instantons are of the form $\partial_{\rho}^l\delta_{\rho = 0}\d^4X$. This is of dimension $l-4$.  We conclude that we can build a point-like measure of the correct dimension as long as $4-bk\ge 0$.

In the more general situation where we have both fermions and scalars, we need to consider the modified $\beta$-function coefficient $\tilde{b}$ with the correction from fermion zero modes. We have shown that, if $\tilde{b}\le 4$, there are infinitely many instanton counter-terms, where the $k$\textsuperscript{th} counter-term corresponds to a $\delta$-function on the point-like instantons of charge $k$.  This makes the theory non-renormalizable.  

Theories which are perturbatively renormalizable ($b \ge 0$) but non-perturbatively non-renormalizable ($\tilde{b} \le 4$) are rare.  As we have mentioned this is not possible in a theory of just gluons and fermions; scalar fields are required.  Further one can have at most $4$ flavours of fermion, since fermions contribute with opposite signs to $b$ and to $\tilde{b}$.


\section{Discussion}

The problem of computing gluon scattering on strong self-dual gauge backgrounds has garnered considerable attention in recent years, for example, tree MHV gluon amplitudes have been computed on self-dual radiative backgrounds \cite{Adamo:2020yzi} and dyons \cite{Garner:2023izn,Adamo:2024xpc,Garner:2024tis}. (See also \cite{Bogna:2023bbd,Adamo:2025vzv} for related works.)

It's natural to ask whether the techniques employed in this paper can be used to compute loop-level amplitudes of self-dual gauge theory on gauge instantons, or on more general self-dual backgrounds.  We foresee two potential challenges.  The first is the issue of anomalies: cancellation of the twistorial gauge anomaly via the Green-Schwarz mechanism follows from taking $N_f=N_c$, and unlike in the case of a flavour background the trace-ordered amplitude is not independent of $N_c$.  Fortunately for $\fg = \fsl(2),\fsl(3)$ this condition is lifted.

In scenarios where the Green-Schwarz mechanism applies, the one-loop all-plus amplitude equates to the integral of a tree-level form factor in the background coupled theory.  The second challenge is evaluating this form factor.  One approach is the chiral algebra bootstrap \cite{Costello:2022wso}, although this requires knowledge of the background deformed chiral algebra.  Of course, even having determined the amplitude on a generic gauge instanton the moduli space integral remains.

Our methods apply somewhat more directly to the computation of all-plus gluon amplitudes on self-dual Einstein backgrounds.  The relevant 't Hooft anomaly and Green-Schwarz mechanism has been analysed in \cite{Bittleston:2022nfr}, reducing the amplitude to an integrated tree-level form factor on the curved geometry.  Again, the difficulty lies in evaluating this form factor.


\begin{acknowledgments}
We thank N. Paquette and A. Strominger for helpful conversations. The authors gratefully acknowledge the support of the Simons Collaboration on Celestial Holography. Research at Perimeter Institute is supported by the Government of Canada through Industry Canada and by the Province of Ontario through the Ministry of Research and Innovation.
\end{acknowledgments}


\bibliography{main}

\begin{thebibliography}{45}%
\makeatletter
\providecommand \@ifxundefined [1]{%
 \@ifx{#1\undefined}
}%
\providecommand \@ifnum [1]{%
 \ifnum #1\expandafter \@firstoftwo
 \else \expandafter \@secondoftwo
 \fi
}%
\providecommand \@ifx [1]{%
 \ifx #1\expandafter \@firstoftwo
 \else \expandafter \@secondoftwo
 \fi
}%
\providecommand \natexlab [1]{#1}%
\providecommand \enquote  [1]{``#1''}%
\providecommand \bibnamefont  [1]{#1}%
\providecommand \bibfnamefont [1]{#1}%
\providecommand \citenamefont [1]{#1}%
\providecommand \href@noop [0]{\@secondoftwo}%
\providecommand \href [0]{\begingroup \@sanitize@url \@href}%
\providecommand \@href[1]{\@@startlink{#1}\@@href}%
\providecommand \@@href[1]{\endgroup#1\@@endlink}%
\providecommand \@sanitize@url [0]{\catcode `\\12\catcode `\$12\catcode `\&12\catcode `\#12\catcode `\^12\catcode `\_12\catcode `\%12\relax}%
\providecommand \@@startlink[1]{}%
\providecommand \@@endlink[0]{}%
\providecommand \url  [0]{\begingroup\@sanitize@url \@url }%
\providecommand \@url [1]{\endgroup\@href {#1}{\urlprefix }}%
\providecommand \urlprefix  [0]{URL }%
\providecommand \Eprint [0]{\href }%
\providecommand \doibase [0]{http://dx.doi.org/}%
\providecommand \selectlanguage [0]{\@gobble}%
\providecommand \bibinfo  [0]{\@secondoftwo}%
\providecommand \bibfield  [0]{\@secondoftwo}%
\providecommand \translation [1]{[#1]}%
\providecommand \BibitemOpen [0]{}%
\providecommand \bibitemStop [0]{}%
\providecommand \bibitemNoStop [0]{.\EOS\space}%
\providecommand \EOS [0]{\spacefactor3000\relax}%
\providecommand \BibitemShut  [1]{\csname bibitem#1\endcsname}%
\let\auto@bib@innerbib\@empty
\bibitem [{\citenamefont {Bern}\ \emph {et~al.}(1994)\citenamefont {Bern}, \citenamefont {Chalmers}, \citenamefont {Dixon},\ and\ \citenamefont {Kosower}}]{Bern:1993qk}%
  \BibitemOpen
  \bibfield  {author} {\bibinfo {author} {\bibfnamefont {Z.}~\bibnamefont {Bern}}, \bibinfo {author} {\bibfnamefont {G.}~\bibnamefont {Chalmers}}, \bibinfo {author} {\bibfnamefont {L.~J.}\ \bibnamefont {Dixon}}, \ and\ \bibinfo {author} {\bibfnamefont {D.~A.}\ \bibnamefont {Kosower}},\ }\href {\doibase 10.1103/PhysRevLett.72.2134} {\bibfield  {journal} {\bibinfo  {journal} {Phys. Rev. Lett.}\ }\textbf {\bibinfo {volume} {72}},\ \bibinfo {pages} {2134} (\bibinfo {year} {1994})},\ \Eprint {http://arxiv.org/abs/hep-ph/9312333} {arXiv:hep-ph/9312333} \BibitemShut {NoStop}%
\bibitem [{\citenamefont {Bern}\ \emph {et~al.}(1993)\citenamefont {Bern}, \citenamefont {Dixon},\ and\ \citenamefont {Kosower}}]{Bern:1993sx}%
  \BibitemOpen
  \bibfield  {author} {\bibinfo {author} {\bibfnamefont {Z.}~\bibnamefont {Bern}}, \bibinfo {author} {\bibfnamefont {L.~J.}\ \bibnamefont {Dixon}}, \ and\ \bibinfo {author} {\bibfnamefont {D.~A.}\ \bibnamefont {Kosower}},\ }in\ \href@noop {} {\emph {\bibinfo {booktitle} {{International Conference on Strings 93}}}}\ (\bibinfo {year} {1993})\ \Eprint {http://arxiv.org/abs/hep-th/9311026} {arXiv:hep-th/9311026} \BibitemShut {NoStop}%
\bibitem [{\citenamefont {Mahlon}(1994)}]{Mahlon:1993si}%
  \BibitemOpen
  \bibfield  {author} {\bibinfo {author} {\bibfnamefont {G.}~\bibnamefont {Mahlon}},\ }\href {\doibase 10.1103/PhysRevD.49.4438} {\bibfield  {journal} {\bibinfo  {journal} {Phys. Rev. D}\ }\textbf {\bibinfo {volume} {49}},\ \bibinfo {pages} {4438} (\bibinfo {year} {1994})},\ \Eprint {http://arxiv.org/abs/hep-ph/9312276} {arXiv:hep-ph/9312276} \BibitemShut {NoStop}%
\bibitem [{Note1()}]{Note1}%
  \BibitemOpen
  \bibinfo {note} {We could also resolve the zero mode issue by simultaneously scattering $2kN_c$ right-handed fermions. Tree level amplitudes of this type have previously been investigated in \cite {HawkingYM,Ringwald:1989ee}.}\BibitemShut {Stop}%
\bibitem [{\citenamefont {Chalmers}\ and\ \citenamefont {Siegel}(1996)}]{Chalmers:1996rq}%
  \BibitemOpen
  \bibfield  {author} {\bibinfo {author} {\bibfnamefont {G.}~\bibnamefont {Chalmers}}\ and\ \bibinfo {author} {\bibfnamefont {W.}~\bibnamefont {Siegel}},\ }\href {\doibase 10.1103/PhysRevD.54.7628} {\bibfield  {journal} {\bibinfo  {journal} {Phys. Rev. D}\ }\textbf {\bibinfo {volume} {54}},\ \bibinfo {pages} {7628} (\bibinfo {year} {1996})},\ \Eprint {http://arxiv.org/abs/hep-th/9606061} {arXiv:hep-th/9606061} \BibitemShut {NoStop}%
\bibitem [{Note2()}]{Note2}%
  \BibitemOpen
  \bibinfo {note} {At least for generic external kinematics. There are non-vanishing distributional terms \cite {Guevara:2026qzd}.}\BibitemShut {Stop}%
\bibitem [{\citenamefont {Ward}(1977)}]{Ward:1977ta}%
  \BibitemOpen
  \bibfield  {author} {\bibinfo {author} {\bibfnamefont {R.~S.}\ \bibnamefont {Ward}},\ }\href {\doibase 10.1016/0375-9601(77)90842-8} {\bibfield  {journal} {\bibinfo  {journal} {Phys. Lett. A}\ }\textbf {\bibinfo {volume} {61}},\ \bibinfo {pages} {81} (\bibinfo {year} {1977})}\BibitemShut {NoStop}%
\bibitem [{\citenamefont {Mason}(2005)}]{Mason:2005zm}%
  \BibitemOpen
  \bibfield  {author} {\bibinfo {author} {\bibfnamefont {L.~J.}\ \bibnamefont {Mason}},\ }\href {\doibase 10.1088/1126-6708/2005/10/009} {\bibfield  {journal} {\bibinfo  {journal} {JHEP}\ }\textbf {\bibinfo {volume} {10}},\ \bibinfo {pages} {009} (\bibinfo {year} {2005})},\ \Eprint {http://arxiv.org/abs/hep-th/0507269} {arXiv:hep-th/0507269} \BibitemShut {NoStop}%
\bibitem [{\citenamefont {Boels}\ \emph {et~al.}(2007)\citenamefont {Boels}, \citenamefont {Mason},\ and\ \citenamefont {Skinner}}]{Boels:2006ir}%
  \BibitemOpen
  \bibfield  {author} {\bibinfo {author} {\bibfnamefont {R.}~\bibnamefont {Boels}}, \bibinfo {author} {\bibfnamefont {L.~J.}\ \bibnamefont {Mason}}, \ and\ \bibinfo {author} {\bibfnamefont {D.}~\bibnamefont {Skinner}},\ }\href {\doibase 10.1088/1126-6708/2007/02/014} {\bibfield  {journal} {\bibinfo  {journal} {JHEP}\ }\textbf {\bibinfo {volume} {02}},\ \bibinfo {pages} {014} (\bibinfo {year} {2007})},\ \Eprint {http://arxiv.org/abs/hep-th/0604040} {arXiv:hep-th/0604040} \BibitemShut {NoStop}%
\bibitem [{\citenamefont {Costello}(2021)}]{Costello:2021bah}%
  \BibitemOpen
  \bibfield  {author} {\bibinfo {author} {\bibfnamefont {K.~J.}\ \bibnamefont {Costello}},\ }\href@noop {} {\enquote {\bibinfo {title} {{Quantizing Local Holomorphic Field Theories on Twistor Space}},}\ } (\bibinfo {year} {2021}),\ \bibinfo {note} {arxiv.org/abs/2111.08879},\ \Eprint {http://arxiv.org/abs/2111.08879} {arXiv:2111.08879 [hep-th]} \BibitemShut {NoStop}%
\bibitem [{\citenamefont {Bardeen}(1996)}]{Bardeen:1995gk}%
  \BibitemOpen
  \bibfield  {author} {\bibinfo {author} {\bibfnamefont {W.~A.}\ \bibnamefont {Bardeen}},\ }\href {\doibase 10.1143/PTPS.123.1} {\bibfield  {journal} {\bibinfo  {journal} {Prog. Theor. Phys. Suppl.}\ }\textbf {\bibinfo {volume} {123}},\ \bibinfo {pages} {1} (\bibinfo {year} {1996})}\BibitemShut {NoStop}%
\bibitem [{\citenamefont {Costello}\ and\ \citenamefont {Paquette}(2022)}]{Costello:2022wso}%
  \BibitemOpen
  \bibfield  {author} {\bibinfo {author} {\bibfnamefont {K.}~\bibnamefont {Costello}}\ and\ \bibinfo {author} {\bibfnamefont {N.~M.}\ \bibnamefont {Paquette}},\ }\href {\doibase 10.1007/JHEP10(2022)193} {\bibfield  {journal} {\bibinfo  {journal} {JHEP}\ }\textbf {\bibinfo {volume} {10}},\ \bibinfo {pages} {193} (\bibinfo {year} {2022})},\ \Eprint {http://arxiv.org/abs/2201.02595} {arXiv:2201.02595 [hep-th]} \BibitemShut {NoStop}%
\bibitem [{\citenamefont {Dixon}\ and\ \citenamefont {Morales}(2024)}]{Dixon:2024mzh}%
  \BibitemOpen
  \bibfield  {author} {\bibinfo {author} {\bibfnamefont {L.~J.}\ \bibnamefont {Dixon}}\ and\ \bibinfo {author} {\bibfnamefont {A.}~\bibnamefont {Morales}},\ }\href {\doibase 10.1007/JHEP08(2024)129} {\bibfield  {journal} {\bibinfo  {journal} {JHEP}\ }\textbf {\bibinfo {volume} {08}},\ \bibinfo {pages} {129} (\bibinfo {year} {2024})},\ \Eprint {http://arxiv.org/abs/2407.13967} {arXiv:2407.13967 [hep-th]} \BibitemShut {NoStop}%
\bibitem [{Note3()}]{Note3}%
  \BibitemOpen
  \bibinfo {note} {Here the disconnected amplitude does not include diagrams with loops of $\protect \mathrm {SU}(N_c)$ gluons. Their contribution is given by the standard formula \cite {Bern:1993sx}.}\BibitemShut {Stop}%
\bibitem [{\citenamefont {Low}(1958)}]{Low:1958sn}%
  \BibitemOpen
  \bibfield  {author} {\bibinfo {author} {\bibfnamefont {F.~E.}\ \bibnamefont {Low}},\ }\href {\doibase 10.1103/PhysRev.110.974} {\bibfield  {journal} {\bibinfo  {journal} {Phys. Rev.}\ }\textbf {\bibinfo {volume} {110}},\ \bibinfo {pages} {974} (\bibinfo {year} {1958})}\BibitemShut {NoStop}%
\bibitem [{\citenamefont {Weinberg}(1965)}]{Weinberg:1965nx}%
  \BibitemOpen
  \bibfield  {author} {\bibinfo {author} {\bibfnamefont {S.}~\bibnamefont {Weinberg}},\ }\href {\doibase 10.1103/PhysRev.140.B516} {\bibfield  {journal} {\bibinfo  {journal} {Phys. Rev.}\ }\textbf {\bibinfo {volume} {140}},\ \bibinfo {pages} {B516} (\bibinfo {year} {1965})}\BibitemShut {NoStop}%
\bibitem [{\citenamefont {Bassetto}\ \emph {et~al.}(1983)\citenamefont {Bassetto}, \citenamefont {Ciafaloni},\ and\ \citenamefont {Marchesini}}]{Bassetto:1983mvz}%
  \BibitemOpen
  \bibfield  {author} {\bibinfo {author} {\bibfnamefont {A.}~\bibnamefont {Bassetto}}, \bibinfo {author} {\bibfnamefont {M.}~\bibnamefont {Ciafaloni}}, \ and\ \bibinfo {author} {\bibfnamefont {G.}~\bibnamefont {Marchesini}},\ }\href {\doibase 10.1016/0370-1573(83)90083-2} {\bibfield  {journal} {\bibinfo  {journal} {Phys. Rept.}\ }\textbf {\bibinfo {volume} {100}},\ \bibinfo {pages} {201} (\bibinfo {year} {1983})}\BibitemShut {NoStop}%
\bibitem [{\citenamefont {Bern}\ \emph {et~al.}(2014)\citenamefont {Bern}, \citenamefont {Davies},\ and\ \citenamefont {Nohle}}]{Bern:2014oka}%
  \BibitemOpen
  \bibfield  {author} {\bibinfo {author} {\bibfnamefont {Z.}~\bibnamefont {Bern}}, \bibinfo {author} {\bibfnamefont {S.}~\bibnamefont {Davies}}, \ and\ \bibinfo {author} {\bibfnamefont {J.}~\bibnamefont {Nohle}},\ }\href {\doibase 10.1103/PhysRevD.90.085015} {\bibfield  {journal} {\bibinfo  {journal} {Phys. Rev. D}\ }\textbf {\bibinfo {volume} {90}},\ \bibinfo {pages} {085015} (\bibinfo {year} {2014})},\ \Eprint {http://arxiv.org/abs/1405.1015} {arXiv:1405.1015 [hep-th]} \BibitemShut {NoStop}%
\bibitem [{\citenamefont {He}\ \emph {et~al.}(2016)\citenamefont {He}, \citenamefont {Mitra},\ and\ \citenamefont {Strominger}}]{He:2015zea}%
  \BibitemOpen
  \bibfield  {author} {\bibinfo {author} {\bibfnamefont {T.}~\bibnamefont {He}}, \bibinfo {author} {\bibfnamefont {P.}~\bibnamefont {Mitra}}, \ and\ \bibinfo {author} {\bibfnamefont {A.}~\bibnamefont {Strominger}},\ }\href {\doibase 10.1007/JHEP10(2016)137} {\bibfield  {journal} {\bibinfo  {journal} {JHEP}\ }\textbf {\bibinfo {volume} {10}},\ \bibinfo {pages} {137} (\bibinfo {year} {2016})},\ \Eprint {http://arxiv.org/abs/1503.02663} {arXiv:1503.02663 [hep-th]} \BibitemShut {NoStop}%
\bibitem [{\citenamefont {'t~Hooft}(1976)}]{tHooft:1976snw}%
  \BibitemOpen
  \bibfield  {author} {\bibinfo {author} {\bibfnamefont {G.}~\bibnamefont {'t~Hooft}},\ }\href {\doibase 10.1103/PhysRevD.14.3432} {\bibfield  {journal} {\bibinfo  {journal} {Phys. Rev. D}\ }\textbf {\bibinfo {volume} {14}},\ \bibinfo {pages} {3432} (\bibinfo {year} {1976})},\ \bibinfo {note} {[Erratum: Phys.Rev.D 18, 2199 (1978)]}\BibitemShut {NoStop}%
\bibitem [{\citenamefont {Vainshtein}\ \emph {et~al.}(1982)\citenamefont {Vainshtein}, \citenamefont {Zakharov}, \citenamefont {Novikov},\ and\ \citenamefont {Shifman}}]{Vainshtein:1981wh}%
  \BibitemOpen
  \bibfield  {author} {\bibinfo {author} {\bibfnamefont {A.~I.}\ \bibnamefont {Vainshtein}}, \bibinfo {author} {\bibfnamefont {V.~I.}\ \bibnamefont {Zakharov}}, \bibinfo {author} {\bibfnamefont {V.~A.}\ \bibnamefont {Novikov}}, \ and\ \bibinfo {author} {\bibfnamefont {M.~A.}\ \bibnamefont {Shifman}},\ }\href {\doibase 10.1070/PU1982v025n04ABEH004533} {\bibfield  {journal} {\bibinfo  {journal} {Sov. Phys. Usp.}\ }\textbf {\bibinfo {volume} {25}},\ \bibinfo {pages} {195} (\bibinfo {year} {1982})}\BibitemShut {NoStop}%
\bibitem [{Note4()}]{Note4}%
  \BibitemOpen
  \bibinfo {note} {We are treating the space of fermions as complex, not real, vector space, in which the path integral is taken over a contour. This is natural because instantons are Euclidean and Euclidean spinors $\psi _\alpha ,\protect \tilde \psi _{\protect \dot \alpha }$ are complex fields.}\BibitemShut {Stop}%
\bibitem [{Note5()}]{Note5}%
  \BibitemOpen
  \bibinfo {note} {At the level of the integrand over instanton moduli space, multi-trace amplitudes are products of single-trace amplitudes, but this is no longer true when we integrate.}\BibitemShut {Stop}%
\bibitem [{\citenamefont {Osborn}(1981)}]{Osborn:1981yf}%
  \BibitemOpen
  \bibfield  {author} {\bibinfo {author} {\bibfnamefont {H.}~\bibnamefont {Osborn}},\ }\href {\doibase 10.1016/0003-4916(81)90159-7} {\bibfield  {journal} {\bibinfo  {journal} {Annals Phys.}\ }\textbf {\bibinfo {volume} {135}},\ \bibinfo {pages} {373} (\bibinfo {year} {1981})}\BibitemShut {NoStop}%
\bibitem [{\citenamefont {Adamo}\ \emph {et~al.}(2023)\citenamefont {Adamo}, \citenamefont {Mason},\ and\ \citenamefont {Sharma}}]{Adamo:2020yzi}%
  \BibitemOpen
  \bibfield  {author} {\bibinfo {author} {\bibfnamefont {T.}~\bibnamefont {Adamo}}, \bibinfo {author} {\bibfnamefont {L.}~\bibnamefont {Mason}}, \ and\ \bibinfo {author} {\bibfnamefont {A.}~\bibnamefont {Sharma}},\ }\href {\doibase 10.1007/s00220-022-04582-9} {\bibfield  {journal} {\bibinfo  {journal} {Commun. Math. Phys.}\ }\textbf {\bibinfo {volume} {399}},\ \bibinfo {pages} {1731} (\bibinfo {year} {2023})},\ \Eprint {http://arxiv.org/abs/2010.14996} {arXiv:2010.14996 [hep-th]} \BibitemShut {NoStop}%
\bibitem [{\citenamefont {Garner}\ and\ \citenamefont {Paquette}(2023)}]{Garner:2023izn}%
  \BibitemOpen
  \bibfield  {author} {\bibinfo {author} {\bibfnamefont {N.}~\bibnamefont {Garner}}\ and\ \bibinfo {author} {\bibfnamefont {N.~M.}\ \bibnamefont {Paquette}},\ }\href {\doibase 10.1007/JHEP08(2023)088} {\bibfield  {journal} {\bibinfo  {journal} {JHEP}\ }\textbf {\bibinfo {volume} {08}},\ \bibinfo {pages} {088} (\bibinfo {year} {2023})},\ \Eprint {http://arxiv.org/abs/2305.00049} {arXiv:2305.00049 [hep-th]} \BibitemShut {NoStop}%
\bibitem [{\citenamefont {Adamo}\ \emph {et~al.}(2025)\citenamefont {Adamo}, \citenamefont {Bogna}, \citenamefont {Mason},\ and\ \citenamefont {Sharma}}]{Adamo:2024xpc}%
  \BibitemOpen
  \bibfield  {author} {\bibinfo {author} {\bibfnamefont {T.}~\bibnamefont {Adamo}}, \bibinfo {author} {\bibfnamefont {G.}~\bibnamefont {Bogna}}, \bibinfo {author} {\bibfnamefont {L.}~\bibnamefont {Mason}}, \ and\ \bibinfo {author} {\bibfnamefont {A.}~\bibnamefont {Sharma}},\ }\href {\doibase 10.1007/s11005-025-01907-2} {\bibfield  {journal} {\bibinfo  {journal} {Lett. Math. Phys.}\ }\textbf {\bibinfo {volume} {115}},\ \bibinfo {pages} {18} (\bibinfo {year} {2025})},\ \Eprint {http://arxiv.org/abs/2406.09165} {arXiv:2406.09165 [hep-th]} \BibitemShut {NoStop}%
\bibitem [{\citenamefont {Garner}\ and\ \citenamefont {Paquette}(2025)}]{Garner:2024tis}%
  \BibitemOpen
  \bibfield  {author} {\bibinfo {author} {\bibfnamefont {N.}~\bibnamefont {Garner}}\ and\ \bibinfo {author} {\bibfnamefont {N.~M.}\ \bibnamefont {Paquette}},\ }\href {\doibase 10.1007/JHEP05(2025)228} {\bibfield  {journal} {\bibinfo  {journal} {JHEP}\ }\textbf {\bibinfo {volume} {05}},\ \bibinfo {pages} {228} (\bibinfo {year} {2025})},\ \Eprint {http://arxiv.org/abs/2408.11092} {arXiv:2408.11092 [hep-th]} \BibitemShut {NoStop}%
\bibitem [{\citenamefont {Bogna}\ and\ \citenamefont {Mason}(2023)}]{Bogna:2023bbd}%
  \BibitemOpen
  \bibfield  {author} {\bibinfo {author} {\bibfnamefont {G.}~\bibnamefont {Bogna}}\ and\ \bibinfo {author} {\bibfnamefont {L.}~\bibnamefont {Mason}},\ }\href {\doibase 10.1007/JHEP08(2023)165} {\bibfield  {journal} {\bibinfo  {journal} {JHEP}\ }\textbf {\bibinfo {volume} {08}},\ \bibinfo {pages} {165} (\bibinfo {year} {2023})},\ \Eprint {http://arxiv.org/abs/2305.07542} {arXiv:2305.07542 [hep-th]} \BibitemShut {NoStop}%
\bibitem [{\citenamefont {Adamo}\ and\ \citenamefont {Ilderton}(2025)}]{Adamo:2025vzv}%
  \BibitemOpen
  \bibfield  {author} {\bibinfo {author} {\bibfnamefont {T.}~\bibnamefont {Adamo}}\ and\ \bibinfo {author} {\bibfnamefont {A.}~\bibnamefont {Ilderton}},\ }\href {\doibase 10.1103/dxl7-m68z} {\bibfield  {journal} {\bibinfo  {journal} {Phys. Rev. D}\ }\textbf {\bibinfo {volume} {111}},\ \bibinfo {pages} {125005} (\bibinfo {year} {2025})},\ \Eprint {http://arxiv.org/abs/2501.06109} {arXiv:2501.06109 [hep-ph]} \BibitemShut {NoStop}%
\bibitem [{\citenamefont {Bittleston}\ \emph {et~al.}(2023)\citenamefont {Bittleston}, \citenamefont {Skinner},\ and\ \citenamefont {Sharma}}]{Bittleston:2022nfr}%
  \BibitemOpen
  \bibfield  {author} {\bibinfo {author} {\bibfnamefont {R.}~\bibnamefont {Bittleston}}, \bibinfo {author} {\bibfnamefont {D.}~\bibnamefont {Skinner}}, \ and\ \bibinfo {author} {\bibfnamefont {A.}~\bibnamefont {Sharma}},\ }\href {\doibase 10.1007/s00220-023-04828-0} {\bibfield  {journal} {\bibinfo  {journal} {Commun. Math. Phys.}\ }\textbf {\bibinfo {volume} {403}},\ \bibinfo {pages} {1543} (\bibinfo {year} {2023})},\ \Eprint {http://arxiv.org/abs/2208.12701} {arXiv:2208.12701 [hep-th]} \BibitemShut {NoStop}%
\bibitem [{\citenamefont {Hawking}\ and\ \citenamefont {Pope}(1979)}]{HawkingYM}%
  \BibitemOpen
  \bibfield  {author} {\bibinfo {author} {\bibfnamefont {S.~W.}\ \bibnamefont {Hawking}}\ and\ \bibinfo {author} {\bibfnamefont {C.~N.}\ \bibnamefont {Pope}},\ }\href {\doibase 10.1016/0550-3213(79)90128-7} {\bibfield  {journal} {\bibinfo  {journal} {Nucl. Phys. B}\ }\textbf {\bibinfo {volume} {161}},\ \bibinfo {pages} {93} (\bibinfo {year} {1979})}\BibitemShut {NoStop}%
\bibitem [{\citenamefont {Ringwald}(1990)}]{Ringwald:1989ee}%
  \BibitemOpen
  \bibfield  {author} {\bibinfo {author} {\bibfnamefont {A.}~\bibnamefont {Ringwald}},\ }\href {\doibase 10.1016/0550-3213(90)90300-3} {\bibfield  {journal} {\bibinfo  {journal} {Nucl. Phys. B}\ }\textbf {\bibinfo {volume} {330}},\ \bibinfo {pages} {1} (\bibinfo {year} {1990})}\BibitemShut {NoStop}%
\bibitem [{\citenamefont {Guevara}\ \emph {et~al.}(2026)\citenamefont {Guevara}, \citenamefont {Lupsasca}, \citenamefont {Skinner}, \citenamefont {Strominger},\ and\ \citenamefont {Weil}}]{Guevara:2026qzd}%
  \BibitemOpen
  \bibfield  {author} {\bibinfo {author} {\bibfnamefont {A.}~\bibnamefont {Guevara}}, \bibinfo {author} {\bibfnamefont {A.}~\bibnamefont {Lupsasca}}, \bibinfo {author} {\bibfnamefont {D.}~\bibnamefont {Skinner}}, \bibinfo {author} {\bibfnamefont {A.}~\bibnamefont {Strominger}}, \ and\ \bibinfo {author} {\bibfnamefont {K.}~\bibnamefont {Weil}},\ }\href@noop {} {\enquote {\bibinfo {title} {{Single-Minus Gluon Tree Amplitudes are Nonzero}},}\ } (\bibinfo {year} {2026}),\ \bibinfo {note} {arxiv.org/abs/2602.12176},\ \Eprint {http://arxiv.org/abs/2602.12176} {arXiv:2602.12176 [hep-th]} \BibitemShut {NoStop}%
\bibitem [{\citenamefont {Hirzebruch}\ \emph {et~al.}(1966)\citenamefont {Hirzebruch}, \citenamefont {Borel},\ and\ \citenamefont {Schwarzenberger}}]{Hirzebruch:1966to}%
  \BibitemOpen
  \bibfield  {author} {\bibinfo {author} {\bibfnamefont {F.}~\bibnamefont {Hirzebruch}}, \bibinfo {author} {\bibfnamefont {A.}~\bibnamefont {Borel}}, \ and\ \bibinfo {author} {\bibfnamefont {R.}~\bibnamefont {Schwarzenberger}},\ }\href@noop {} {\emph {\bibinfo {title} {Topological Methods in Algebraic Geometry}}},\ Vol.\ \bibinfo {volume} {175}\ (\bibinfo  {publisher} {Springer Berlin-Heidelberg-New York},\ \bibinfo {year} {1966})\BibitemShut {NoStop}%
\bibitem [{Note6()}]{Note6}%
  \BibitemOpen
  \bibinfo {note} {Field theories on twistor space are analytically continued theories, whose space of fields is an infinite-dimensional complex manifold but where the path integral is taken over a contour. The real form of the gauge algebra is specified by the choice of contour. Perturbatively, the choice of contour doesn't matter.}\BibitemShut {Stop}%
\bibitem [{Note7()}]{Note7}%
  \BibitemOpen
  \bibinfo {note} {There are also additional gauge transformations $\chi ^\prime $ under which $\protect \mathcal {B}$ transforms as $\delta \protect \mathcal {B}= \protect \bar {\partial }\chi ^\prime + [\protect \mathcal {A},\chi ^\prime ]$. Similarly, the twistor uplifts of fermions and scalars introduced below have their own gauge transformations.}\BibitemShut {Stop}%
\bibitem [{Note8()}]{Note8}%
  \BibitemOpen
  \bibinfo {note} {One can also derive the anomaly from the Grothendieck-Hirzebruch-Riemann-Roch formula \cite {Hirzebruch:1966to}, which tells us that the first Chern class of the determinant line bundle where the partition function lives is $\protect \tfrac {1}{2}\DOTSI \intop \ilimits@ _\protect \PazoBB {PT}\protect \big [\protect \operatorname {Td}\protect \big (T_\protect \PazoBB {PT}\protect \big )\protect \mathrm {Ch}\protect \big (\protect \mathfrak {g}+ \protect \mathscr {O}(-4)\otimes \protect \mathfrak {g}- \protect \mathscr {O}(-1)\otimes R_f - \protect \mathscr {O}(-3)\otimes R^\vee _f + \protect \mathscr {O}(-2)\otimes (R_s\oplus R_s^\vee )\protect \big )\protect \big ]_8$ where the subscript $8$ means we are integrating the $8$-form component. The only non-vanishing contribution in our situation is that from $\protect \mathrm {td}_0$ and $\protect \mathrm {ch}_4$.}\BibitemShut {Stop}%
\bibitem [{\citenamefont {Green}\ and\ \citenamefont {Schwarz}(1984)}]{Green:1984sg}%
  \BibitemOpen
  \bibfield  {author} {\bibinfo {author} {\bibfnamefont {M.~B.}\ \bibnamefont {Green}}\ and\ \bibinfo {author} {\bibfnamefont {J.~H.}\ \bibnamefont {Schwarz}},\ }\href {\doibase 10.1016/0370-2693(84)91565-X} {\bibfield  {journal} {\bibinfo  {journal} {Phys. Lett. B}\ }\textbf {\bibinfo {volume} {149}},\ \bibinfo {pages} {117} (\bibinfo {year} {1984})}\BibitemShut {NoStop}%
\bibitem [{\citenamefont {Mason}(1987)}]{Mason:FAv1}%
  \BibitemOpen
  \bibfield  {author} {\bibinfo {author} {\bibfnamefont {L.}~\bibnamefont {Mason}},\ }\href@noop {} {\bibfield  {journal} {\bibinfo  {journal} {Twistor Newsletter}\ }\textbf {\bibinfo {volume} {23}},\ \bibinfo {pages} {36} (\bibinfo {year} {1987})}\BibitemShut {NoStop}%
\bibitem [{\citenamefont {Dixon}\ \emph {et~al.}(2004)\citenamefont {Dixon}, \citenamefont {Glover},\ and\ \citenamefont {Khoze}}]{Dixon:2004za}%
  \BibitemOpen
  \bibfield  {author} {\bibinfo {author} {\bibfnamefont {L.~J.}\ \bibnamefont {Dixon}}, \bibinfo {author} {\bibfnamefont {E.~W.~N.}\ \bibnamefont {Glover}}, \ and\ \bibinfo {author} {\bibfnamefont {V.~V.}\ \bibnamefont {Khoze}},\ }\href {\doibase 10.1088/1126-6708/2004/12/015} {\bibfield  {journal} {\bibinfo  {journal} {JHEP}\ }\textbf {\bibinfo {volume} {12}},\ \bibinfo {pages} {015} (\bibinfo {year} {2004})},\ \Eprint {http://arxiv.org/abs/hep-th/0411092} {arXiv:hep-th/0411092} \BibitemShut {NoStop}%
\bibitem [{\citenamefont {Atiyah}\ \emph {et~al.}(1978)\citenamefont {Atiyah}, \citenamefont {Hitchin}, \citenamefont {Drinfeld},\ and\ \citenamefont {Manin}}]{Atiyah:1978ri}%
  \BibitemOpen
  \bibfield  {author} {\bibinfo {author} {\bibfnamefont {M.~F.}\ \bibnamefont {Atiyah}}, \bibinfo {author} {\bibfnamefont {N.~J.}\ \bibnamefont {Hitchin}}, \bibinfo {author} {\bibfnamefont {V.~G.}\ \bibnamefont {Drinfeld}}, \ and\ \bibinfo {author} {\bibfnamefont {Y.~I.}\ \bibnamefont {Manin}},\ }\href {\doibase 10.1016/0375-9601(78)90141-X} {\bibfield  {journal} {\bibinfo  {journal} {Phys. Lett. A}\ }\textbf {\bibinfo {volume} {65}},\ \bibinfo {pages} {185} (\bibinfo {year} {1978})}\BibitemShut {NoStop}%
\bibitem [{\citenamefont {Horrocks}(1964)}]{Horrocks:1964vb}%
  \BibitemOpen
  \bibfield  {author} {\bibinfo {author} {\bibfnamefont {G.}~\bibnamefont {Horrocks}},\ }\href {\doibase https://doi.org/10.1112/plms/s3-14.4.689} {\bibfield  {journal} {\bibinfo  {journal} {Proceedings of the London Mathematical Society}\ }\textbf {\bibinfo {volume} {s3-14}},\ \bibinfo {pages} {689} (\bibinfo {year} {1964})}\BibitemShut {NoStop}%
\bibitem [{\citenamefont {Witten}(1996)}]{Witten:1995gx}%
  \BibitemOpen
  \bibfield  {author} {\bibinfo {author} {\bibfnamefont {E.}~\bibnamefont {Witten}},\ }\href {\doibase 10.1016/0550-3213(95)00625-7} {\bibfield  {journal} {\bibinfo  {journal} {Nucl. Phys. B}\ }\textbf {\bibinfo {volume} {460}},\ \bibinfo {pages} {541} (\bibinfo {year} {1996})},\ \Eprint {http://arxiv.org/abs/hep-th/9511030} {arXiv:hep-th/9511030} \BibitemShut {NoStop}%
\bibitem [{\citenamefont {Douglas}(1999)}]{Douglas:1995bn}%
  \BibitemOpen
  \bibfield  {author} {\bibinfo {author} {\bibfnamefont {M.~R.}\ \bibnamefont {Douglas}},\ }\href@noop {} {\bibfield  {journal} {\bibinfo  {journal} {NATO Sci. Ser. C}\ }\textbf {\bibinfo {volume} {520}},\ \bibinfo {pages} {267} (\bibinfo {year} {1999})},\ \Eprint {http://arxiv.org/abs/hep-th/9512077} {arXiv:hep-th/9512077} \BibitemShut {NoStop}%
\end{thebibliography}%
\bibliographystyle{apsrev4-1}
\clearpage


\widetext
\clearpage
\begin{center}
\textbf{\large Supplemental Materials}
\end{center}


\section{Proof of Theorem \ref{thm:instanton_amplitudes}} \label{sec:thm_proof}

Here we detail the proof of Theorem \ref{thm:instanton_amplitudes}, which is based on twistor methods.

 
\subsection{Theories on Twistor Space} \label{supp:twistors}

Classically, self-dual gauge theory arises from a holomorphic theory on twistor space \cite{Ward:1977ta,Mason:2005zm,Boels:2006ir}.  Twistor space is the complex manifold
\be \PT = \Oo(1)\oplus\Oo(1)\to\CP^1\,. \ee
As a real manifold, it is diffeomorphic to $\R^4 \times S^2$.  The complex structure of twistor space encodes the standard conformal structure of $\R^4$.

The twistor uplift of self-dual Yang-Mills theory with real gauge algebra $\mf{g}_\R$ is holomorphic BF theory with complex gauge algebra $\mf{g} = \mf{g}_\C$ \footnote{Field theories on twistor space are analytically continued theories, whose space of fields is an infinite-dimensional complex manifold but where the path integral is taken over a contour.  The real form of the gauge algebra is specified by the choice of contour.  Perturbatively, the choice of contour doesn't matter.}.  The fields of holomorphic BF theory consists of a $(0,1)$-form gauge field
\be \cA\in\Omega^{0,1}(\PT,\mf{g}) \ee
and a $(3,1)$-form Lagrange multiplier field
\be \cB\in\Omega^{3,1}(\PT,\mf{g}) \ee
with action
\be \label{eqn:hBF} \frac{\im}{2\pi}\int_\PT \op{tr}(\cB\wedge F(\cA))\,. \ee
$\cA$ is subject to gauge symmetries parametrised infinitesimally by $\Omega^{0,0}(\PT,\mf{g})$.  Since $\cA$ is only a $(0,1)$-form, the gauge transformations involve the Dolbeault instead of the de Rham operator, and take the form
\be \delta\cA = \dbar\chi + [\cA,\chi]\,. \ee
$\cB$ transforms in the adjoint representation of the gauge transformations generated by $\chi$ \footnote{There are also additional gauge transformations $\chi^\prime$ under which $\cB$ transforms as $\delta\cB = \dbar\chi^\prime + [\cA,\chi^\prime]$. Similarly, the twistor uplifts of fermions and scalars introduced below have their own gauge transformations.}.

If we include Weyl fermions in some representation $R_f$, then we also adjoin fields
\be \widetilde\Psi_+ \in \Pi\Omega^{0,1}(\PT, \Oo(-1) \otimes R_f)\,,\qquad \Psi_- \in \Pi\Omega^{0,1}(\PT, \Oo(-3) \otimes R^\vee_f) \ee
with action
\be \label{eqn:twistor_Weyls} \frac{\im}{2\pi}\int_\PT\la\Psi_-,\dbar_\cA\widetilde\Psi_+\ra_{R_f}\,. \ee
To include complex scalars in the representation $R_s$ we adjoin
\be (\widetilde\Phi,\Phi)\in \Omega^{0,1}(\PT,\Oo(-2)\otimes(R_s\oplus R_s^\vee)) \ee
with Lagrangian
\be \label{eqn:twistor_Higgs} \frac{\im}{2\pi}\int_\PT\la\Phi,\dbar_\cA\widetilde\Phi\ra_{R_s}\,. \ee

Compactifying the theory \eqref{eqn:hBF} together with fermions \eqref{eqn:twistor_Weyls} and scalars \eqref{eqn:twistor_Higgs} along the fibration $\PT \to \R^4$ yields self-dual gauge theory with the action \eqref{eqn:sdgaugetheory}.


\subsection{Twistorial Anomalies}

At loop-level, this theory suffers from gauge anomalies on twistor space \cite{Costello:2021bah}, even when we assume that $R_f$ is a real representation so that there is no chiral anomaly on space-time. 

The twistor space anomaly comes from a box diagram on twistor space, which fails to be gauge invariant \footnote{One can also derive the anomaly from the Grothendieck-Hirzebruch-Riemann-Roch formula \cite{Hirzebruch:1966to}, which tells us that the first Chern class of the determinant line bundle where the partition function lives is
$\half\int_\PT\big[\op{Td}\big(T_\PT\big)\Cha\big(\fg + \Oo(-4)\otimes\fg - \Oo(-1)\otimes R_f - \Oo(-3)\otimes R^\vee_f + \Oo(-2)\otimes (R_s\oplus R_s^\vee)\big)\big]_8$ where the subscript $8$ means we are integrating the $8$-form component.  The only non-vanishing contribution in our situation is that from $\todd_0$ and $\cha_4$.}. Its gauge variation is 
\be \frac{1}{3!}\bigg(\frac{\im}{2\pi}\bigg)^3\int_\PT\op{tr}_{\mf{g} - R_f+R_s}\big(\chi(\p\cA)^3\big)\,. \label{eqn:anomaly}
\ee
For certain gauge algebras $\mf{g}$ and matter representations $R_f,R_s$ this expression vanishes, or more generally can be cancelled via a Green-Schwarz mechanism \cite{Costello:2021bah}.  When the twistorial anomaly vanishes the corresponding space-time theory has vanishing loop-level amplitudes, at least for generic kinematics, allowing us to infer information about scattering from anomalies on twistor space \cite{Costello:2022wso}.

Let us briefly review why there is no scattering. Using coordinates $v^\da,z$ on the $\Oo(1)$ fibres and $\CP^1$ base respectively, a plane wave state with momentum given by a pair of spinors $\lt_\da$ and $\lambda = (1,z_0)$ lifts to the state 
\be \label{eqn:scattering_state} e^{v^\da\lt_\da}\bar\delta_{z = z_0} \ee
on twistor space.

If we scatter $n$ plane wave states with generic kinematics, they are all supported at different values of $z$.  If the theory on twistor space is local and holomorphic, then it does not depend on the metric on the $z$-sphere; this is just a gauge choice.  We can take this to be as large as we like, in which case the states are very far apart from each other and there is no scattering.


\subsection{Zero Modes on Twistor Space}

There is a crucial caveat to this argument when the fermionic fields have zero modes.  In this scenario we need to separately perform the integral over the zero modes, and since their twistor representatives are not localised at a single value of $z$ they will generically overlap with the uplifts of scattering states \eqref{eqn:scattering_state}.

We can overcome this issue by introducing a chiral mass term for the fermions, eliminating the zero modes and effectively turning them into parity reversed scalars.  We saw how this works on space-time.  On twistor space the chiral mass term takes the form
\be \label{eq:twistor-chiral-mass} \int_\PT H\wedge (\widetilde\Psi_+,\widetilde\Psi_+)_{R_f} \ee
where $H\in\Omega^{1,1}(\PT)$ represents the hyperplane class on $\PT$ and $(~\,,~)_{R_f}$ is a non-degenerate, symmetric and invariant bilinear on $R_f$ encoding the mass.  This deforms the fermion bundle to the extension
\be 0\to \Oo(-3)\otimes R_f^\vee\to E\to \Oo(-1)\otimes R_f\to 0 \ee
by the non-trivial element $H\otimes(~\,,~)_{R_f}\in\mrm{Ext}^1(\Oo(-1),\Oo(-3))\otimes\mrm{Sym}^2(R_f^\vee)$.  $E$ is isomorphic to
\be \Oo(-2)\otimes(R_f^\vee\oplus R_f^\vee) \ee
and comes equipped with the pairing $(~\,,~)_{R_f}^{-1}$.  We recognise this as the twistor uplift of a parity reversed complex scalar in the representation $R_f$.

The chiral mass term violates scaling symmetry of the theory, but this can be rectified by twisting the action of space-time scalings on the fermions by an axial symmetry preserving the pairing $(~\,,~)_{R_f}$.  This regrading modifies the scale anomaly, due to the familiar four-dimensional axial anomaly.  


\subsection{Anomaly Cancellation by the Green-Schwarz Mechanism}

Let us take the complexified gauge algebra to be $\mf{g} = \mf{sl}(N_c)$ with $N_f$ fermion flavours. This means that the matter representation is
\begin{equation}
R_f = \C^{N_f} \otimes (F \oplus F^\vee)
\end{equation}
where $F$ is the fundamental representation of $\mf{sl}(N_c)$.  We will take $R_s = 0$, but as explained above the fermions may be interpreted as parity reversed scalars so our computation extends trivially to a non-trivial $R_s$.  In this case, the anomaly on twistor space is
\be \frac{1}{3!}\bigg(\frac{\im}{2\pi}\bigg)^3\int_\PT\op{tr}_{F\otimes F^\vee - \C^{N_f}\otimes(F\oplus F^\vee)}\big(\chi(\p\cA)^3\big)\,.    
\ee
We can rewrite this in terms of single- and double-traces in the fundamental, as
\be \label{eqn:trace_identity} \op{tr}_{F\otimes F^\vee - \C^{N_f}\otimes(F\oplus F^\vee)}\big(\chi(\p\cA)^3\big) = 6\op{tr}\big(\chi\p\cA\big)\op{tr}\big((\partial\cA)^2\big) + 2(N_c - N_f)\op{tr}\big(\chi(\p\cA)^3\big)\,. \ee
In particular, if $N_f = N_c$ the single-trace part of the anomaly vanishes.
    
The double trace term can be cancelled by a Green-Schwarz mechanism \cite{Green:1984sg,Costello:2021bah}.  To do this, we introduce a $(2,1)$-form field $\eta$ on twistor space, constrained to satisfy $\p\eta = 0$.  The kinetic term coincides with that of Kodaira-Spencer theory
\begin{equation} 
\frac{\im}{4\pi}\int_\PT\p^{-1}\eta \dbar\eta\,.
\end{equation}
This is coupled to the gauge field $\cA$ by a term
\begin{equation} 
\mu\bigg(\frac{\im}{4\pi}\bigg)\int_\PT\eta\op{tr}(\cA\p\cA)\,. \label{eqn:gscoupling} 
\end{equation}
There is a tree-level anomaly in the diagram involving the exchange of a single $\eta$ which cancels the double trace one-loop gauge theory anomaly if we tune
\be \mu^2 = \frac{\lambda^2_{\fg,R_f}}{3}\bigg(\frac{1}{2\pi}\bigg)^2\,. \ee
Here $\lambda^2_{\fg,R_f}$ is the coefficient of the double trace term in equation \eqref{eqn:trace_identity}, so for our chosen matter is $6$.

On space-time the field $\eta$ corresponds \cite{Mason:FAv1,Costello:2021bah} to a dimension zero scalar field $\rho$, with kinetic term
\be
\frac{1}{2}\int_{\R^4}\d^4x\,(\tri\rho)^2
\ee
coupled to self-dual gauge theory by the axion term
\be
\mu\int_{\R^4}\rho\op{tr}\big(F(A)\wedge F(A)\big) \,.
\ee

We conclude that the double trace one-loop amplitude in self-dual gauge theory with $N_f = N_c$ coupled to a dimension zero axion is equal to the tree-level amplitude coming from a single axion exchange.  One can verify this explicitly \cite{Dixon:2024mzh}.  Our goal is to apply the same method to compute scattering in the presence of instantons.


\subsection{Non-Trivial Backgrounds for Flavour Symmetry}

We are interested in the presence of a non-trivial holomorphic $\mf{sl}(N_f)$ bundle on twistor space. We write the corresponding rank $N_f$ bundle as $V$, and the gauge field for this bundle as $\alpha \in \Omega^{0,1}(\PT,\mf{sl}(N_f))$. The matter bundle is
\begin{equation}
    R_f = V \otimes F \oplus V^\vee \otimes F^\vee
\end{equation}
where $F$ as above is the fundamental of $\mf{sl}(N_c)$. 
The action is
\begin{equation} 
\frac{\im}{2\pi}\int_\PT \op{tr}(\mc{B}\wedge F(\mc{A})) + \la\Psi_-,\dbar_{\mc{A}+\alpha}\widetilde\Psi_+\ra_{R_f}\,. 
\end{equation} 
The presence of the background gauge field $\alpha$ introduces a mixed gauge-flavour anomaly on twistor space associated to a box diagram whose gauge variation is
\bea
&- \bigg(\frac{\im}{2\pi}\bigg)^3\int_\PT \op{tr}_F\big(\chi\p\mc{A}\big)\op{tr}_V\big((\partial\alpha)^2\big)\,.
\eea
The minus sign here appears because fermions flow through the loop.  A factor of $\half$ from the symmetry of the diagram cancels against a factor of $2$ since the fermions are Diracs.

There's also a pure flavour 't Hooft anomaly which is not relevant for our analysis. The mixed gauge-flavour anomaly can be cancelled by the same Green-Schwarz mechanism as before, where we add the source term
\begin{equation}
- \mu\bigg(\frac{\im}{4\pi}\bigg)\int_\PT \eta \op{tr}_V ( \alpha \partial \alpha) \label{eqn:source_twistor} 
\end{equation}
to the Lagrangian.  The exchange of a single $\eta$ field where one side is coupled to $\op{tr}_F(\mc{A}\partial\mc{A})$ and the other side to $\op{tr}_V (\alpha \partial\alpha)$ cancels the mixed gauge-flavour anomaly.

After applying the Penrose transform, the new coupling \eqref{eqn:source_twistor} gives a term where the axion field $\rho$ is coupled to the field strength $F(a)$ of the background $\mf{su}(N_f)$ gauge field.  We have determined a twistorial theory on space-time depending on the background gauge field $a$:
\bea
\int_{\R^4}\op{tr}(B \wedge F(A)_-) + \la\psi_-,\slashed{D}_{A + a} \tilde\psi_+\ra_{R_f} + \frac{1}{2}\d^4x\,(\tri\rho)^2 + \mu\rho\big(\op{tr}(F(A)^2) - \op{tr}(F(a)^2)\big)\,.
\eea
The coupling of $\rho$ to $\op{tr}(F(a)^2)$ can be viewed as giving a source for $\rho$.

After appropriately regulating the zero modes, this theory has no scattering.  Therefore, the one-loop amplitude generated by the bifundamental fermion loop in the presence of the background gauge field $a$ is cancelled by a tree-level diagram involving a single $\rho$ exchange. The axion exchange must connect the source term $\rho\op{tr}(F(a)^2)$ with the terms coupling $\rho$ to $A$.


\subsection{Computing Amplitudes using Axion Exchange}

The source term is
\be 8\pi^2\mu\int_{\R^4}\rho(x)\mc{D}(a,x) \ee
where $\mc{D}(a,x)$ is the normalized instanton density.  Fixing the value of $x$, the amplitudes arising from axion exchange are form factors of the operator $\rho(x)\mc{D}(a,x)$ in the theory where we've dropped all dependence on the background field, but retained the $\rho$ field. 

The form factor where we insert the operator $\rho$ at the origin and at $x$ differ by an overall factor of $e^{\im P \cdot x}$, where $P$ is the total momentum of the external particles we are scattering.  

The form factor at $x = 0$ is very simple: the trace-ordered form factor is simply the Parke-Taylor denominator
\begin{equation}
\ip{\rho(0) \mid 1^+ \dots n^+} = \frac{\mu}{2\ip{12} \dots \ip{n1} }\,.
\end{equation}
This follows from the result of \cite{Dixon:2004za} which shows that the trace-ordered form factor of the operator $\op{tr}(F(A)^2)$ is 
\begin{equation} 
\ip{\op{tr}\big(F(A)^2\big) \mid 1^+ \dots n^+ } = \frac{ (P \cdot P)^2 } {2\ip{12}\dots\ip{n1 }} 
\end{equation}
where $P$ is the total momentum of the $n$ external particles.  The $\rho-\rho$ propagator cancels the factor of $(P \cdot P)^2$. 

We conclude that the form factor of $\rho(x)\mc{D}(a,x)$ is
\bea
&\ip{\rho(x)\mc{D}(a,x) \mid 1^+ \dots n^+ } = \frac{\mu}{2\ip{12} \dots \ip{n1} }e^{\im P \cdot x}\mc{D}(a,x)\,.
\eea

Integrating this over $x$, and remembering to reverse the sign, gives rise to the one-loop trace-ordered amplitude in the presence of the background field $a$:
\bea \label{eqn:bg_scattering}
\mc{A}^\text{trace-ordered}(1^+,\dots,n^+;a) &= - 4\pi^2\mu^2 \frac{\what{\mc{D}}(a,P)}{\ip{12} \dots \ip{n1}} = - \frac{2\what{\mc{D}}(a,P)}{\ip{12} \dots \ip{n1}}
\eea
where $\what{\mc{D}}(a,P)$ is the Fourier transform of $\mc{D}(a,x)$ and $P$ is the total momentum.  We have checked the normalization of the amplitude in Appendix \ref{supp:normalization} by comparing to the amplitude in the presence of a singular instanton.

We obtained this formula under the assumption that $N_f=N_c$, which was necessary to ensure vanishing of the gauge twistorial anomaly. However, it's clear from the relevant space-time Feynman diagrams that the result is independent of $N_c$. Our formula is therefore valid in the case that $N_f\neq N_c$, though this will change the perturbative contributions to the one-loop all-plus amplitudes from the trivial flavour bundle.

Equation \eqref{eqn:bg_scattering} is the formula for the connected, trace-ordered amplitude. In the totally soft regime the correlators of the centrally extended $S$-algebra compute the disconnected, non-trace-ordered amplitude. It is easy to see that in this limit equation \eqref{eqn:bg_scattering} coincides with the trace-ordered correlator of the centrally-extended $S$-algebra for $\gSU(N_c)$ at level $2k$.  The algebraic manipulations which pass us from connected, trace-ordered expressions to disconnected expressions are the same for the chiral algebra and for the amplitudes.  Equation \ref{eqn:disconnected} follows.


\section{Normalizing the Scattering} \label{supp:normalization}

To normalize the scattering amplitude in an instanton background, we will consider a very simple case: a point-like instanton of charge $k$.  Instantons can be realised by the ADHM construction, which, of course, arises from twistor theory.  We will use this to determine the normalization.  

As above, let $V$ denote the holomorphic bundle on $\PT$ associated to the background field for flavour symmetry.  Let us assume that $V$ is given by the ADHM construction \cite{Atiyah:1978ri}.  This means that it is the cohomology of a Horrocks \cite{Horrocks:1964vb} monad, which is a complex of vector bundles of the form
\begin{equation}
\begin{tikzcd}[row sep = huge]
-1 &  & \Oo(-1)^k \arrow{dl}[swap]{v^\da\op{Id} - X^\da(z)} \arrow{d}{J(z)} \\
0 & \Oo^k \otimes S_+ \arrow[swap]{dr} {v_\da\op{Id} - X_\da(z)} & \Oo^{N_f} \arrow{d}{I(z)} \\
1  &  & \Oo(1)^k
\end{tikzcd}
\end{equation}
Here $S_+$ denotes the two-dimensional spin representation of $\gSO(4)$ with dotted spinor indices.  The $k\times k$ matrices $X^\da(z)$, $k\times N_f$ matrix $I(z)$, and $N_f\times k$ matrix $J(z)$ are polynomials of degree $1$ in $z$.  The coefficients of these polynomials give the usual ADHM matrices.

Let us call this complex of bundles $\mc{V}_{X,I,J}$.  In the background of an ADHM instanton, the matter fields can be taken to live in the parity reversed Dolbeault complex of $\PT$ with values in 
\be (F \otimes \mc{V}_{X,I,J} \oplus F^\vee \otimes \mc{V}^\vee_{X,I,J}  ) \otimes \Oo(-1)[1] \oplus (F \otimes \mc{V}_{X,I,J} \oplus F^\vee \otimes \mc{V}^\vee_{X,I,J}) \otimes \Oo(-3)[1]\,. \ee
(Here, we are writing fields in the BV formalism, where they live in the whole Dolbeault complex; see \cite{Costello:2021bah} for background.  The symbol $[1]$ indicates a shift of cohomological degree by $1$.)  Working with $\mc{V}_{X,I,J}$ is equivalent to working with its cohomology bundle $V_{X,I,J} = H^0(\mc{V}_{X,I,J})$.

The normalizing factor of the $S$-matrix in the instanton background is independent of the chosen instanton, and so can be computed in the limit when we set $X^\da\to0$, $I \to 0$, $J \to 0$.  In this limit, the complex $\mc{V}_{X,I,J}$ has cohomology
\be H^0(\mc{V}_{X,I,J}) = \Oo^{N_f}\,,\qquad H^1(\mc{V}_{X,I,J}) = \Oo_{\CP^1}(1)^k \ee
where $\Oo_{\CP^1}$ is the structure sheaf of the $\CP^1$ at $v^\da = 0$ where the degenerate instanton is centred. (In string theory terms, we have shrunk the instanton to zero and pulled off a $D1$ brane \cite{Witten:1995gx,Douglas:1995bn}.)

In this degenerate limit, the flavour bundle is a sum of a trivial bundle and the sheaf $\Oo_{\CP^1}(1)^k[-1]$.  The scattering amplitude is a sum of contributions from these two sheaves.  We subtract the perturbative contribution coming from the trivial bundle.  The interesting contribution is that from $\Oo_{\CP^1}(1)^k[-1]$.  

Dropping the contribution of the trivial bundle, we find that the fermions on $\PT$ have localized to live on a $\CP^1$, and have become
\be \Pi\Omega^{0,\ast}\big(\CP^1,(F\oplus F^\vee)\otimes\Oo^k\oplus(F^\vee\oplus F)\otimes\Oo(-2)^k\big)\,. \ee
This describes a system of free chiral fermions $(\tilde\chi^i,\tilde\chi_j),(\chi_k,\chi^l)$ living on the twistor line of the origin and valued in $(F \oplus F^\vee)^k$.  The $\tilde\chi$ fields have spin $0$ whereas the $\chi$ fields have spin $1$.  The bulk holomorphic BF theory couples to this system by the Kac-Moody current, which has level $2k$.

Since the $\tilde\chi$ fields have spin $0$, they have zero modes. These correspond to the zero modes of the fermions $\tilde\psi_+$ in the presence of a flavour instanton.  Strictly speaking, the correlation functions are zero because of the zero mode integral.  However, the Kac-Moody central charge appears at the level of the OPEs, which is sufficient to fix the normalization.

The calculation goes through  through largely unmodified for fundamental complex scalars, except that the Kac-Moody level is now negative.


\section{Analytic Continuation of the Fourier Transform of the Instanton Density} \label{supp:analytic}

In this section we will study the analytic properties  of the Fourier transformed instanton density $\what{\mc{D}}(a,P)$ associated to an instanton $a$ coming from the ADHM construction. We will show that $\what{\mc{D}}(a,P)$ is analytic on $\C^4$ with no poles, but with branch cuts along the light cone $P \cdot P = 0$. 

An ADHM instanton determined by $I(z),J(z),X^\da(z)$ obeying the ADHM equation has density \cite{Osborn:1981yf}
\be \mc{D}(I,J,X;x) = \frac{1}{16\pi^2}\triangle^2\mrm{tr}_{\C^k}\log f\,, \ee
where
\be \label{eqn:ADHM_mx} f = \frac{2}{J_\al I^\al + (x - X)^2}\,. \ee
Here we've expanded $I(z) = I^\al\lambda_\al$, $J(z) = J^\al\lambda_\al$ and $X^\da(z) = X^{\da\al}\lambda_\al$ for $\lambda = (1,z)$.  We've also abbreviated $(x - X)^2 = (x_{\da\al} - X_{\da\al})(x^{\da\al} - X^{\da\al})$.

The ADHM matrices are invariant under combined Hermitian and Euclidean spinor conjugation.  This ensures that the denominator of the ADHM matrix \eqref{eqn:ADHM_mx} is non-negative.

The Fourier transform
\be \what{\mc{D}}(I,J,X;p) = \frac{1}{16\pi^2}\int_{\R^4}\d^4x\,e^{\im p\cdot x} \triangle^2\mrm{tr}_{\C^k}\log f \label{eqn:fourier} \ee
is certainly well-defined for Euclidean momenta $p$, because $\triangle^2\op{tr}_{\C^k}\log f$ is square-integrable.  To extend it to complex momenta, we take the integral \eqref{eqn:fourier} defining the Fourier transform, and perform the integral over some contour given by $x$ where, for large $x$, $p \cdot x$ is real.  For small $x$, we can take $x$ to be Euclidean so that we avoid any poles in $\det f$.  We need to show that if $p$ is not null, then the integral converges absolutely with this choice of contour. 

Now, 
\be \log f = \log 2 - \log (x \cdot x) - \log\left(1 - \frac{2X\cdot x}{x\cdot x} + \frac{J_\al I^\al + X^2}{x\cdot x}\right) \ee
Since $\tri^2 \log (x \cdot x)$ is a $\delta$-function, this term does not contribute to convergence of the integral.  

If $p$ is not null, then $x$ lives on a contour which, for large $x$, is a copy of $\R^4$ in $\C^4$ on which $p \cdot x$ is real.  The restriction of the holomorphic metric on $\C^4$ to the contour $\R^4$ gives it a complex-valued metric. If $p$ is not null, we can always choose a contour for which this metric has no null vectors.  To see this, we note that by a transformation in $\gSO_4(\C)$ we can assume that $p$ is a vector of the form $(p_1,0,0,0)$.  For $x$  far away from zero, we can take the contour to consist of vectors of the form $\bar{p}_1 (x_1 , x_2,x_3,x_4)$ where the $x_i$ are real.   The metric on this contour differs by a phase from the Euclidean metric, and so has no null vectors. 

For large $\abs{x \cdot x}$, we can expand
\be \log\left(  1 - \frac{ 2 X \cdot x } { x \cdot x} + \frac{J_\al I^\al + X^2 }{ x \cdot x}\right) = - \frac{2X \cdot x}{x \cdot x} + \frac{J_\al I^\al + X^2}{ x \cdot x} - \frac{2(X\cdot x)^2}{(x\cdot x)^2} + \dots\,. \label{eqn:expansion} \ee
This expansion converges.  Applying $\tri^2$, we find an expansion which starts with a term that goes like $(x \cdot x)^{-3}$.  (This is because the terms that appear to go like $r^{-4}$ or like $r^{-5}$ are both zero. For example, the $r^{-5}$ term is  $\tri^2 ( x / (x \cdot x)  )$ is a derivative of a $\delta$-function at $x = 0$). We conclude that $\tri^2 \log f$ goes like $(x \cdot x)^{-3}$ for large $x$, so that the integral defining the Fourier transform is absolutely convergent on this contour.

If $p \cdot p =0$ this argument does not apply, and we find a branch cut because of the change in contour as we move around the null cone.

Next let us determine the expansion of $\what{\mc{D}}(a,p)$ near $p\cdot p = 0$. We will show that
\be
   \what{\mc{D}}(a,p) = F(p) +   G(p) (p\cdot p)^2  \log (p \cdot p) 
\ee
where $F(p)$,  $G(p)$ are  entire holomorphic functions of $p$. 

To see this, we write the integral defining the Fourier transform as a sum of an integral over $\norm{x} \le R$ and an integral over $\norm{x} \ge R$.  The integral over small $\norm{x}$ yields an entire holomorphic function of $p$.  For large $ \norm{x}$, we have the convergent expansion \eqref{eqn:expansion}.  Each term in the expansion is a matrix product of some number of copies of $(X\cdot x)/\norm{x}^2$ with some number of copies of $(X^2 + J_\al I^\al)/\norm{x}^2$.  We therefore need to consider the Fourier transform of
\be \label{eqn:expansionterm} \frac{x^{\mu_1}\dots x^{\mu_k}}{\norm{x}^{2(m+k)}}\,. \ee
If $m > 0$ this is a sum of derivatives of $\norm{x}^{-2(m+l)}$ for $0 \le l \le \lfloor k/2 \rfloor$. If $m = 0$, we also find derivatives of $\log \norm{x}^2$.  Any such expression is either a derivative of $\log \norm{x}^2$, or of $\norm{x}^{-4}$. 

Recalling that we are applying $\tri^2$ before we Fourier transform, we see that those terms which are derivatives of $\log \norm{x}^2$ only contribute polynomials in $p$. This is because $\tri^2 \log \norm{x}^2$ is a $\delta$-function.  The Fourier transform of $\norm{x}^{-4}$ is $\log (p \cdot p)$, so that -- incorporating the contribution from $\tri^2$ and other derivatives -- terms which are derivatives of $\norm{x}^{-4}$ contribute $G(p)(p \cdot p)^2 \log (p \cdot p)$ for some polynomial $p$.

 In the one instanton case, the Fourier transform can be evaluated directly. 
 Writing $J_\al I^\al = \rho^2$, $R^2 = \rho^2 + (x-X)^2$ we have
\be \triangle\log f = - \p_{\da\al}\bigg(\frac{2(x-X)^{\da\al}}{R^2}\bigg) = - \frac{4}{R^2} - \frac{4\rho^2}{R^4}\,. \ee
Similarly
\bea
&\triangle^2\log f = 8\p_{\da\al}\bigg(\frac{(x-X)^{\da\al}}{R^4} + \frac{2\rho^2(x-X)^{\da\al}}{R^6}\bigg) = 32\bigg(\frac{1}{R^4}+\frac{2\rho^2}{R^6} - \frac{R^2-\rho^2}{R^6} - \frac{3(R^2-\rho^2)\rho^2}{R^8}\bigg) = \frac{96\rho^4}{R^8}\,.
\eea
Then
\be \widehat{\mc{D}}(a,p) = \frac{6\rho^4}{\pi^2}\int_{\R^4}\mathrm{d}^4x\,\frac{e^{\mrm{i}p\cdot x}}{(\rho^2+(x-X)^2)^4}\,. \ee
Fourier transforms of this type are standard
\bea
&\int_{\R^4}\mrm{d}^4x\,\frac{e^{\mrm{i}p\cdot x}}{(\rho^2+(x-X)^2)^n} = \frac{1}{\Gamma(n)}e^{\im p\cdot X}\int_0^\infty\mrm{d}\mu\,\mu^{n-1}e^{-\mu\rho^2}\int_{\R^4}\d^4x\,e^{-\mu x^2 + \mrm{i}p\cdot x} \\
&= \frac{\pi^2}{\Gamma(n)}e^{\im p\cdot X}\int_0^\infty\d\mu\,\mu^{n-3}e^{-\mu\rho^2 - p^2/4\mu} = \frac{2\pi^2}{\Gamma(n)}e^{\im p\cdot X}\bigg(\frac{|p|}{2\rho}\bigg)^{n-2}K_{|2-n|}(\rho|p|)\,.
\eea


\section{Exponential Decay of Fourier Transformed Instanton Density at Large $\rho$} \label{supp:decay}

The Bessel function $K_2(z)$ decays exponentially for large $z$ (with $\abs{\arg z}\leq3\pi/2)$, 
\be
K_2(z) \sim \sqrt{\frac{\pi}{2z}} e^{-z}\,.
\ee
This means that for real momenta $p$ (or momenta close to real), large instantons of charge $1$ are exponentially suppressed.  We will show a similar result for instantons of higher charge.

We will use the following facts about $\mc{D}(I,J,X;x)$ (denoted $\mc{D}(x)$ from now on):
\begin{enumerate}
\item $\mc{D}(x)$ is a rational function of $x$ with real coefficients.
\item $\mc{D}(x)$ has no poles for $x \in \R^4$. 
\item $\mc{D}(x)$ for large $x$ has a convergent series expansion starting with $r^{-6}$.
\end{enumerate}
By multiplying $\mc{D}(x)$ by a constant, we can ensure that $\abs{\mc{D}(x)} \le 1$ for all $x$.

We let
\be
C = \op{Supp}\{ \eps \ge 0\mid \forall~x, y \in \R^4 \text{ with } \norm{y} = \eps\,,~\abs{\mc{D}(x+ \im y)} \le 2 \}\,.
\ee
First, we will show that $C > 0$.   To see this, we use the fact that has a convergent series expansion starting with $r^{-6}$ in the region $\norm{x} \ge  R$ for some $R$.   Because this series is convergent, it defines an analytic function of $x + \im y$ in some domain. This implies  that for $\norm{x} \ge R$, $\abs{\mc{D}(x+\im y)}$ is bounded above by $2$ as long as $y$ is not too large. 

Next, let us consider performing the Fourier transform of $\mc{D}(x)$ over a contour where $x$ has been shifted by $C \im  p / \norm{p}$. (We will assume for now that the momentum $p$ is real.)

More specifically, we will take the contour given by $x + \im y$ where:
\begin{enumerate}
\item If $\norm{x} \ge R$, for some fixed large $R$, then $y = 0$.
\item If $\norm{x} < R$, then $y =  C  p/\norm{p}$. 
\item If $\norm{x} = R$, then $y = \delta  p/ \norm{p}$ for $0 \le \delta \le C$.
\end{enumerate}
Let us perform the Fourier transform integral
\be
\int \d^4x\,e^{\im p \cdot x} \mc{D}(x)
\ee
over this contour.   We will find bounds for the integral over the three regions.

To determine the bounds, we let $\mc{D}_1(x)= \tri^2 \log (1 + \norm{x}^2 )$ be the instanton density for an instanton of charge $k = 1$ and size $\rho = 1$, and let
\be
D = \op{sup}_x \frac{\mc{D}(x) } {\mc{D}_1(x) }
\ee
be the $L^1$ norm of the ratio $\mc{D}(x) / \mc{D}_1(x)$.  This is finite as both $\mc{D}(x)$ and $\mc{D}_1(x)$ decay like $r^{-6}$ for large $x$. 

For $R$ sufficiently large, $\mc{D}(x)$ behaves like $D \norm{x}^{-6} + \mrm{O}(\norm{x}^{-7})$.  This tells us that the contribution of the integral over the region $\norm{x} \le R$ is bounded above by $D R^{-2} + \mrm{O}(R^{-3})$.  

Again using the fact that $\mc{D}(x)$ admits a convergent expansion in the region $\norm{x} \gg 0$, and bounding the $\norm{x}^{-6}$ term in this expansion by the corresponding term in $\mc{D}_1(x)$, we find that the contribution over the region $\norm{x} = R$ is bounded by a multiple of
\be
\frac{D}{p^2R^6} + \mrm{O}\bigg(\frac{1}{R^7}\bigg)\,.
\ee
For the region $\norm{x} \le R$, $\mc{D}(x + C \im p / \norm{p} )$ is bounded above by $2$.  Therefore the integral over this region is bounded by a multiple of
\begin{equation}
 R^4 e^{- C \norm{p}}\,.
\end{equation}
Putting this all together, we find that the integral is bounded by a sum of multiples of $R^{-2}$, $R^{-6}$ and $R^4 e^{-C \norm{p} }$.  The integral is independent of $R$, so we can take $R = e^{\alpha \norm{p} }$ for $0 < \alpha < C/4$.  This gives a bound which is a sum of $e^{-2 \alpha \norm{p}}$, $e^{-6 \alpha \norm{p}} $ and $e^{(4 \alpha - C) \norm{p} }$.  We can take $\alpha = C/6$, so that the integral is bounded by $e^{-C \norm{p}/3}$.




\end{document}